\documentclass[aip,apl,amsmath,amssymb,reprint,nofootinbib]{revtex4-1}
\pdfoutput=1

\usepackage{graphicx}
\usepackage{dcolumn}
\usepackage{epsfig}
\usepackage{epstopdf}
\usepackage{color}
\usepackage[normalem]{ulem}
\usepackage{color,soul}
\usepackage{amssymb}
\usepackage{ifpdf}

\newcommand*{\SRO} {SrRuO$_{3}$}

\newcommand*{\STO} {SrTiO$_{3}$}

\newcommand*{\BFO} {BiFeO$_{3}$}
\newcommand*{\LSMO}{La$_{0.67}$Sr$_{0.33}$MnO$_{3}$}

\newcommand{\curpath}{figures/}
\graphicspath{\curpath}

\begin{document}
	\preprint{AIP/123-QED}
	%\title[]{Recent advances in large area Pulsed Laser Deposition; epitaxial growth of complex oxides on silicon}
	\title[]{Epitaxial growth of complex oxides on silicon by enhanced surface diffusion in large area pulsed laser deposition.}
	\author{Rik Groenen}
	\affiliation{Twente Solid State Technology B.V., 7500 AG Enschede, The Netherlands}
	\author{Zhaoliang Liao}
	\affiliation{Faculty of Science and Technology and MESA+ Institute for Nanotechnology, University of Twente, 7500 AE Enschede, The Netherlands}
	\author{Nicolas Gauquelin}
	\affiliation{Antwerp University, 2000 Antwerpen, Belgium}
	\author{Roel Hoekstra}
	\author{Bart Spanjer}
	\author{Merle van Gorsel}
	\author{Sam Borkent}
	\affiliation{Twente Solid State Technology B.V., 7500 AG Enschede, The Netherlands}
	\affiliation{Saxion University of Applied Sciences, 7500 KB Enschede, The Netherlands}
	\author{Minh Nguyen}
	\affiliation{Faculty of Science and Technology and MESA+ Institute for Nanotechnology, University of Twente, 7500 AE Enschede, The Netherlands}
	\author{Laura Vargas-LLona}
	\affiliation{Faculty of Science and Technology and MESA+ Institute for Nanotechnology, University of Twente, 7500 AE Enschede, The Netherlands}
	\author{Eddy Rodijk}
	\affiliation{Twente Solid State Technology B.V., 7500 AG Enschede, The Netherlands}
	\author{Cas Damen}
	\affiliation{Twente Solid State Technology B.V., 7500 AG Enschede, The Netherlands}
	\affiliation{Saxion University of Applied Sciences, 7500 KB Enschede, The Netherlands}
	\author{Jo Verbeeck}
	\affiliation{Antwerp University, 2000 Antwerpen, Belgium}
	\author{Gertjan Koster}
	\email{g.koster@utwente.nl}	
	\author{Guus Rijnders}
	\affiliation{Faculty of Science and Technology and MESA+ Institute for Nanotechnology, University of Twente, 7500 AE Enschede, The Netherlands}
	\date{\today}
	
\begin{abstract}

Homogeneous highly epitaxial {\LSMO}(LSMO) thin films have been grown on Yttria-stabilized-Zirconia (YsZ) / CeO$_2$ buffer layers on technological relevant 4'' silicon wafers using a Twente Solid State Technology B.V. (TSST) developed large area Pulsed Laser Deposition (PLD) setup. We study and show the results of the effect of an additional {\SRO} buffer layer on the growth temperature dependent structural and magnetic properties of LSMO films. With the introduction of a thin \SRO\ layer on top of the buffer stack, LSMO films show ferromagnetic behaviour for growth temperatures as low as 250$^{\circ}$C. We suggest that occurrence of epitaxial crystal growth of LSMO at these low growth temperatures can be understood by an improved surface diffusion, which ensures sufficient intermixing of surface species for formation of the correct phase. This intermixing is necessary because the full plume is collected on the 4'' wafer resulting in a compositional varying flux of species on the wafer, in contrast to small scale experiments.

\end{abstract}

\maketitle

%\section{Introduction}
The discovery of colossal magnetoresistance in the perovskite manganites in 1993 resulted in an extensive amount of research, because of the potential for data storage applications. Among the perovskite manganites, LSMO has the largest single electron bandwidth and the highest Curie temperature, making LSMO an interesting material for application in spintronic devices. Examples are magnetic tunnel junctions, Schottky devices, and magnetoelectric devices. All these applications require LSMO thin films where the device performance will mainly depend on the thin film quality. 
Many studies have shown the potential of Pulsed Laser Deposition (PLD) as a suitable deposition technique for the growth of high quality LSMO films.\cite{Boschker2011a} These studies often stress the potential of functional devices using the properties of these high quality LSMO thin films, mostly demonstrated on lab-scale (small) single crystalline substrates using relatively high growth temperatures.

However, functional devices demand the growth of films on substrate materials with sizes suitable for industrial applications, where silicon wafers define the standard of CMOS technology. This will require upscaling of the PLD process to grow films on wafer-sized substrates.  

Two important challenges in thin film growth on silicon wafers using PLD have been the focus of our research in recent years. First, epitaxial growth of oxides on as-received silicon wafers is intrinsically prohibited by an amorphous native silicon dioxide layer, demanding a procedure to first remove this native oxide layer. Growth of YsZ buffer layers on silicon has proven to be an effective method to remove the native silicon dioxide layer to form an epitaxial basis for oxide growth. The zirconium reduces the native oxide, forming an epitaxial relation between YsZ and silicon.\cite{DeCoux2012}$^{,}${\cite{Wang2000}$^{,}${\cite{Aguiar1997} CeO$_2$ is typically used to overcome the large lattice mismatch between YsZ and perovskites\cite{Dekkers2009}, intrinsically also forming a beneficial chemical buffer layer between the reactive zirconium and perovskites. Other alternatives to obtain an epitaxial relation between oxide thin film and silicon have also been explored, for example growing \STO\ on silicon using MBE.\cite{Brooks2009} Secondly, specifically in PLD, the dimensions of the plasma plume are typically much smaller than commonly used industrial wafer sizes, which is solved by scanning the plasma plume over the full wafer area \cite{TSST}. 

One final hurdle for full CMOS process compatibility remains: significantly lower process temperatures are required compared to typical small scale PLD growth experiments. Little research has been carried out focussing on optimisation of growth at lower temperatures in PLD experiments, because generally this is expected to lower the crystallinity.   

In order to enable growth of high quality oxide materials at lower temperatures, one has to somehow compensate for the unavoidable reduced surface diffusivities, typically leading to a higher probability of formation of growth defects including thermodynamically unfavourable phases. Kinetic models state that surface diffusion $D_S$ depends on temperature and activation energy for diffusion, or $D_S \sim exp\big(-\frac{E_A}{{K_b}T}\big)$. Therefore, at a given activation energy, lowering the temperature leads to an exponential decrease of the diffusivity. On the other hand, lowering of the activation energy itself could compensate for the effect of a lower undesired temperature.

Studies have shown the relation between terminating crystal plane, activation energy and therefore surface diffusion \cite{Rijnders2004}. These studies, focussing on the growth of \SRO\ investigated by RHEED, have shown the occurrence of a termination switch during the initial growth of \SRO\ resulting in SrO or A-site terminated growth. This significantly alters the growth kinetics, understood by an improved surface diffusion due to a lower activation energy on specifically the A-site termination. More recent work on BiFeO$_3$ growth on A- and B-site terminated \STO\ substrates also show this drastic change in growth kinetics understood by improved growth kinetics on A-site terminated {\STO}.\cite{Solmaz2016} 

Here we study the effect of an additional \SRO\ buffer layer on the structural and magnetic properties of LSMO thin films grown on optimised YsZ/CeO$_2$ buffer layers on 4'' silicon wafers using scanning PLD. A LSMO growth temperature dependence is shown, where with the introduction of a thin \SRO\ layer in between CeO$_2$ and LSMO, LSMO films show nearly single phase crystallinity and ferromagnetic behaviour for growth temperatures as low as 250$^{\circ}$C. 

%We suggest that occurrence of high quality crystal growth of LSMO at these low growth temperatures, can be understood by an improved surface diffusion induced by the introduced A-site terminated \SRO\ layer. This is necessary to promote species intermixing which ensures the formation of the thermodynamic preferred crystalline phase. This intermixing is necessary because the full plume is collected on the 4'' wafer resulting in a compositional varying flux of species on the wafer. This is in contrast to small scale experiments, where the plume is stationary and only part of the plume is used.

%\section{Experimental methods}
For sample fabrication a specially designed TSST PLD system is used with which is possible to deposit on 4'' wafers. Most importantly, growth parameters can be set to values comparable to a typical small scale PLD system. With a 4'' radiative heater, wafer growth temperatures of up to 900$^{\circ}$C are achieved. 

The most significant difference in design compared to small scale systems concerns the challenge to grow homogeneously on this larger substrate size. The ablation plume is significantly smaller than the 4'' wafer resulting in an inhomogeneous thickness profile when collected statically on substrates larger than 1''. To obtain homogeneous films on larger substrates, the laser spot is scanned over the target, therefore spatially scanning the plume over the 4'' substrate. While scanning the plume in lateral direction the heater rotates to ensure a full coverage of the plume over the full wafer. 
In all experiments presented here, the plume scanning curve, wafer rotation and laser repetition rate are set such that over the full wafer an average growth speed is obtained comparable with small scale experiments of $\sim$2nm/min with a thickness inhomogeneity better then 5\%.   

P-type doped (100) oriented 4'' silicon wafers are used without any specific chemical surface treatment. Buffer layer growth settings are based on optimised small scale experiments. Growth temperature for YsZ and CeO$_2$ is 800$^{\circ}$C. Fluence, spotsize and target-to-substrate distance are respectively 2.2J/cm$^2$, 2.5mm$^2$ and 55mm for all experiments. Growth speeds are determined from preliminary growth experiments and subsequent film thickness measurements with cross-sectional Scanning Electron Microscopy (SEM) and Transmission Electron Microscopy (TEM). First, to promote the silicon native oxide reduction, $\sim$2nm YsZ is grown at a pressure of 0.02mbar argon and a laser repetition rate of 10Hz. Subsequently, a $\sim$100nm thick layer of YsZ is grown at 0.02mbar oxygen and a laser repetition rate of 20Hz. Then a $\sim$8nm thin film of CeO$_2$ is grown in the same pressure conditions at laser repetition rate of 10Hz. A relatively thin layer compared to YsZ is chosen as this CeO$_2$ layer tend to roughen in morphology with increasing thickness. Also, TEM imaging, shown in figure \ref{TEM}, shows that the aimed lattice relaxation of the CeO$_2$ occurs after four unit cells thickness. 

\begin{figure}[h]
    \includegraphics[width=0.4\textwidth]{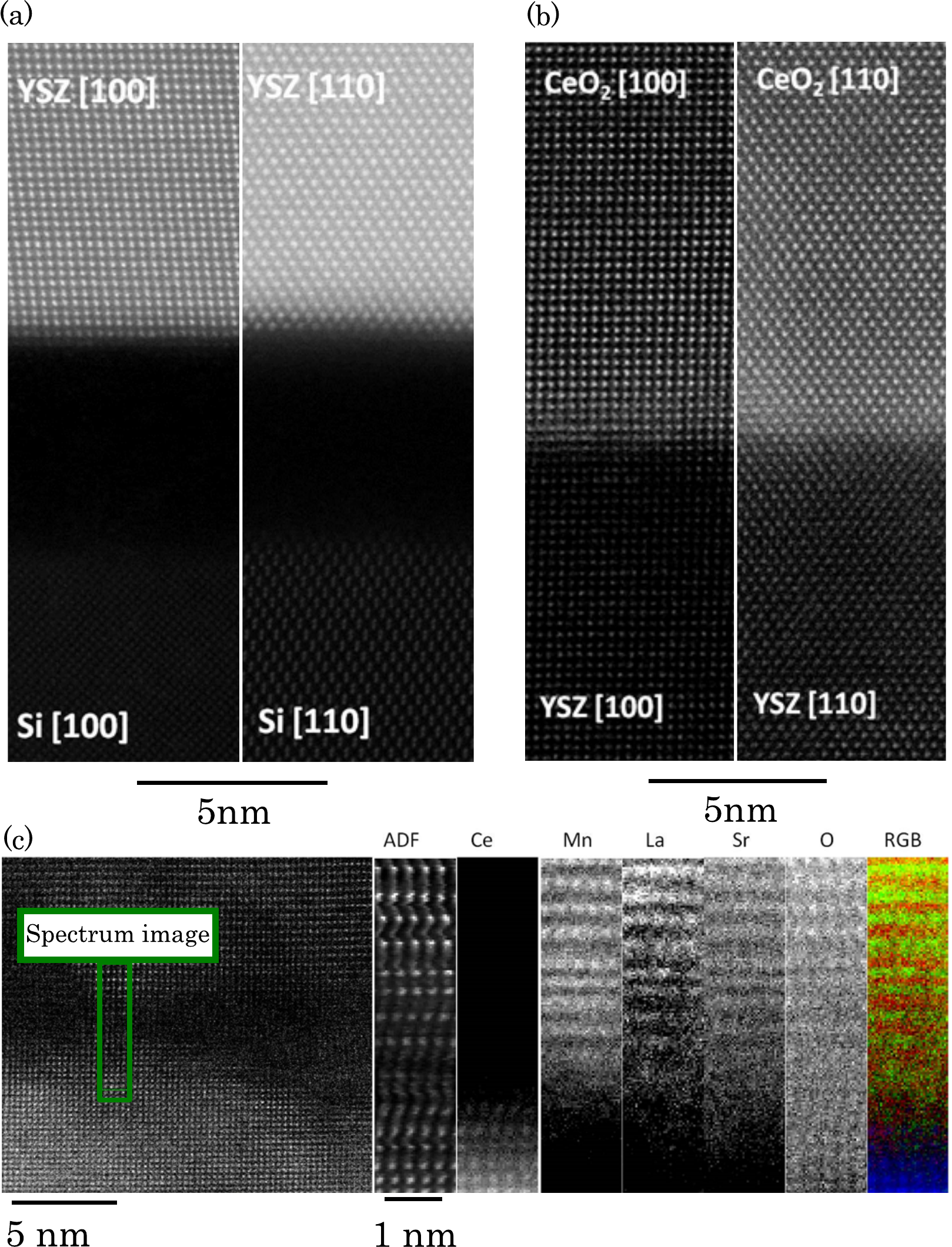}
    \caption{HR-TEM image on the three different interfaces in the full stack. (a) shows the Si // YsZ interface. (b) shows the perfect epitaxial relation between CeO$_2$ and YsZ. (c) EELS imaging shows the element specific sharp interface between LSMO and CeO$_2$.}
       \label{TEM}
\end{figure}

Figure \ref{TEM}(a) shows high resolution High Angle Annular Dark Field (HAADF) images of the Si/YsZ interface. An orientation relationship of Si$<$001$>$YsZ$<$001$>$ is present. A 6nm layer of SiO$_2$ is observed. This indicates that after the growth of YsZ in reduced condition forming an epitaxial relation with the silicon substrate, the top 6nm of silicon below the YsZ re-oxidises. The YsZ/SiO$_2$ interface is sharp with no silicon present in the zirconia region. 

 \begin{figure}[h]
    \includegraphics[width=0.3\textwidth]{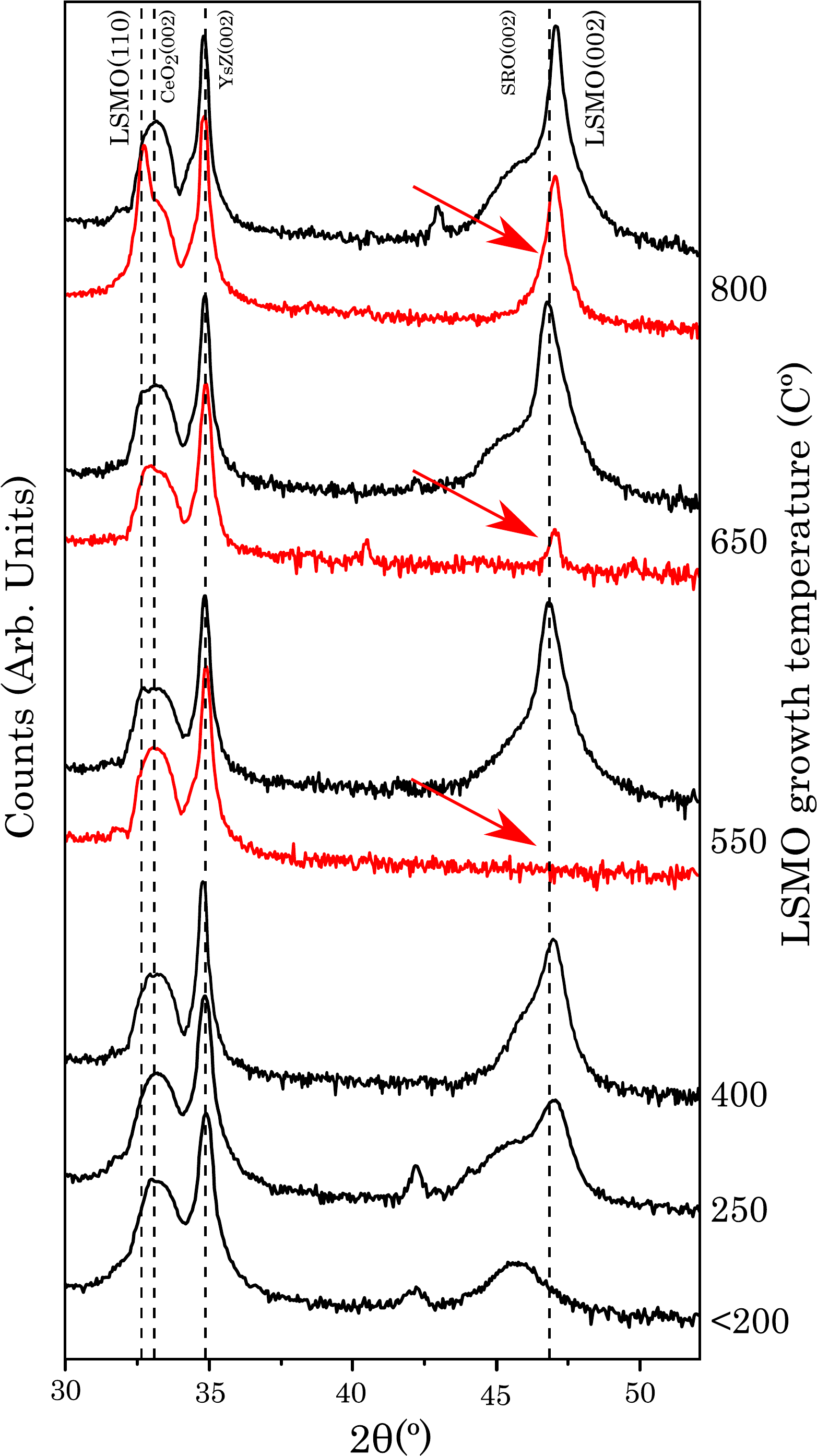}
    \caption{XRD 2$\theta$/$\omega$ symmetrical scans for LSMO films grown on YsZ/CeO$_2$/\SRO\ buffer layers (black) and on YsZ/CeO$_2$ buffer layers (red), without \SRO\. Dashed lines indicates the angle corresponding to bulk c-axis values for respectively cubic fluorite CeO$_2$ (5.41\AA) and YsZ (5.14\AA) and pseudocubic LSMO (3.876\AA).}
    \label{XRD}
\end{figure}

Figure \ref{TEM}(b) shows a very well defined interface between CeO$_2$ and YsZ with an orientation relationship of YsZ$<$001$>$CeO$_2$$<$001$>$. A four unit cell region in the CeO$_2$ is observed with a slightly brighter contrast which is due to strain induced by the 6\% lattice mismatch between the two structures. The discussed stack so far shows a similar structure to what has been published previously \cite{}REF.

Next, for a set of samples, a $\sim$4nm thin \SRO\ film is grown at 1.0*10$^{-6}$mbar and 10Hz. Last, the LSMO film is grown at 0.1mbar oxygen and a laser repetition rate of 20Hz at varying temperatures with a thickness of $\sim$ 100nm. TEM and X-ray Diffraction (XRD) 2$\theta$/$\omega$ symmetrical scan measurements were carried out for structural characterisation. The magnetic properties of LSMO have been investigated with a Quantum Design Vibrating Sample Magnetometer (VSM), with which critical Curie temperature (T$_C$) and the magnetisation of the samples is determined.

%\section{Results and discussion}

Figure \ref{TEM}(c) shows the interface between LSMO and CeO$_2$ with LSMO grown at 800$^{\circ}$C. The interface globally is quite rough but locally sharp nonetheless. The RGB map consists of manganite in red, lanthanum in green and cerium in blue. Note that the thicknesses of the CeO$_2$ and \SRO\ layers in the TEM measurements deviate from the thicknesses used in the remainder of the article. The presence of \SRO\ grown in between the LSMO and CeO$_2$ cannot be confirmed by the EELS measurements. This needs further investigation but should not affect the main conclusions of this article. Based on the TEM observations, for the remainder of the text it was chosen to decrease the CeO$_2$ layer thickness to 20\% suppress to roughening while maintaining the aimed lattice relaxation, whereas the \SRO\ layer was set to double the thickness to ensure full \SRO\ coverage. 
  
Figure \ref{XRD} shows the XRD results for a series of films varying in LSMO growth temperature, all measured at the center of the wafer. In black, measurements are shown for films grown with an \SRO\ layer between CeO$_2$ and LSMO; in red, stacks without this \SRO\ layer. For all films, buffer layers are grown under identical circumstances. Peak reflections have been identified as indicated. For all films, clear single phase c-axis out-of-plane oriented YsZ and CeO$_2$ are observed related to the highly epitaxial relation between the buffer layers and silicon as observed in the TEM measurements. Peak positions correspond to an out-of-plane lattice constant for YsZ and CeO$_2$ of respectively 5.14\AA\ and 5.41\AA\, corresponding to expected bulk values. For the films grown without \SRO\, a strong temperature dependent crystallinity of the LSMO film is observed. At the highest growth temperature of 800$^\circ$C, two clear out-of-plane oriented phases are observed, identified as LSMO(001) and LSMO(110). At lower growth temperature of 650$^\circ$C, a clear decay of these reflections is observed, were at 550$^\circ$C, no crystalline LSMO phases are present. 
For films grown with \SRO\ layer, single phase c-axis oriented LSMO is observed, where unlike films without \SRO\, the observed temperature dependence is lacking; crystalline LSMO phases remain present for these relatively low temperatures. As shown in the bottom scans, films grown at 400$^\circ$C and even as low as 250$^\circ$C still show epitaxial crystalline LSMO growth. LSMO phases and crystallinity disappear for films grown below 200$^\circ$C. 

\begin{center}
\begin{figure}[h]
    \includegraphics[width=0.4\textwidth]{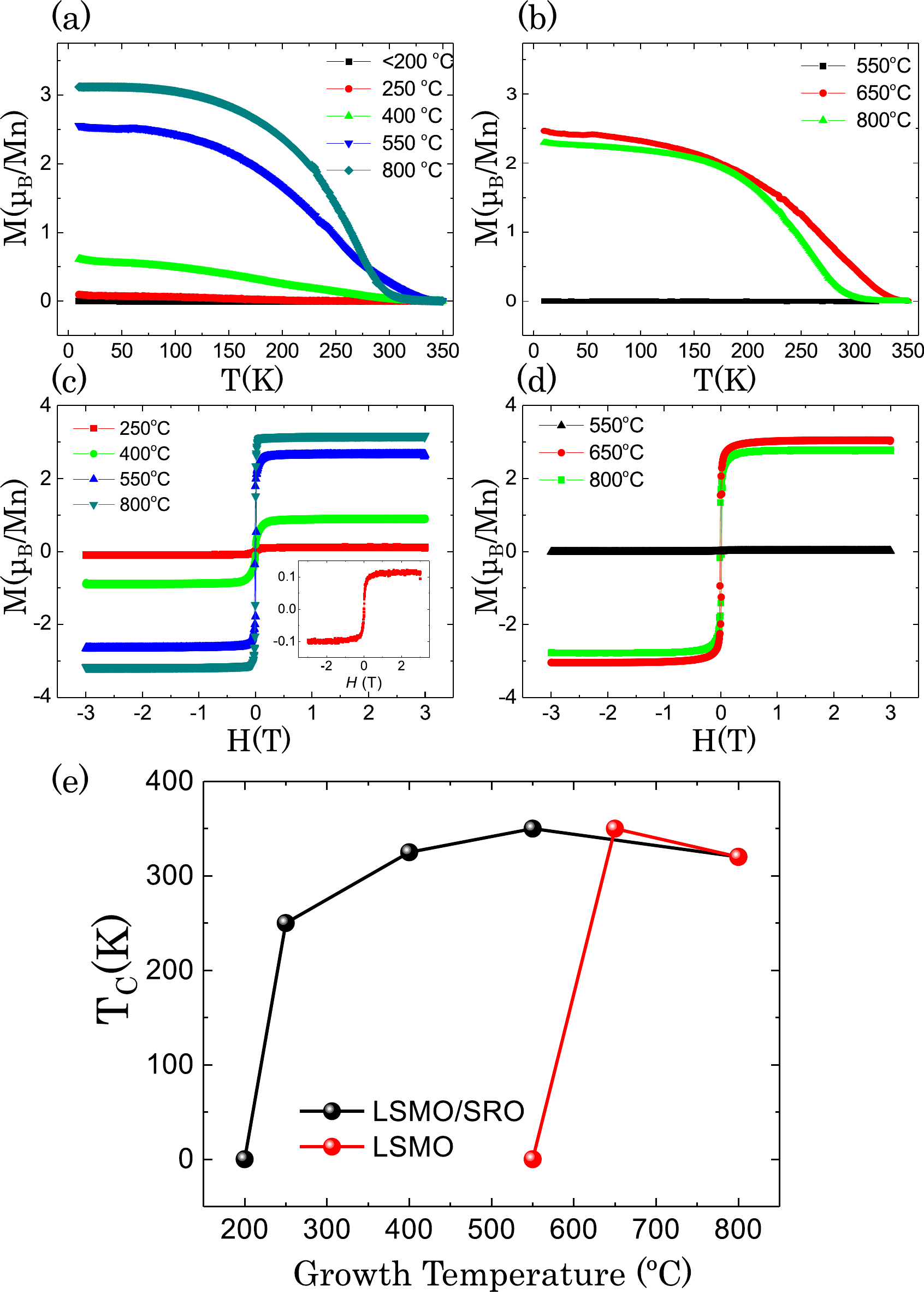}
    \caption{(a-d) Growth temperature dependent magnetization of LSMO films with and without \SRO\ buffer layer in between the LSMO and CeO$_2$ films. (e) shows the Curie temperature (T$_C$) as a function of growth temperature.}
        \label{MagMeasAll}
\end{figure}
\end{center}

Concomitant to different crystallinity, the LSMO films exhibit different magnetisation. As shown in figure \ref{MagMeasAll}(a) showing the temperature dependent magnetisation of \SRO\ buffered LSMO films, the ferromagnetic ordering disappears for amorphous LSMO films grown at low temperatures (T$<$200$^\circ$C). However, as the films become crystalline when the growth temperature is increased above a critical temperature (200$^\circ$C), the films become ferromagnetic and the magnetic moment increases with increasing growth temperature. In contrast, the films without an \SRO\ buffered layer require higher growth temperatures (T$>$550$^\circ$C) to render the films ferromagnetic. As shown in figure \ref{MagMeasAll}(b), a non-buffered film grown at 550$^\circ$C is not magnetic while the films grown at 650$^\circ$C and above are ferromagnetic. Figures \ref{MagMeasAll}(c-d) show the field dependent magnetic moment at 100K of \SRO\ buffered and non-buffered LSMO films. Increased saturated magnetic moments can be observed as the growth temperature is increased. At very low growth temperature of just 250$^\circ$C, for \SRO\ buffered films one can still observe a magnetic moment of 0.1$\mu$B/Mn (see inset of figure \ref{MagMeasAll}(c)).

Figure \ref{MagMeasAll}(e) further shows the Curie temperature (T$_C$) as a function of growth temperature. For the buffered films, the critical temperature for onset of T$_C$ is 250$^\circ$C while for non-buffered film, it is 650$^\circ$C. Note that the thickness used to determine the magnetisation is estimated from cross-sectional SEM measurements by taking an average value. The films grown at higher temperatures were typically the smoothest and therefore magnetisation could be determined more accurately.

We suggest that occurrence of high quality crystal growth of LSMO at these low growth temperatures when grown on a \SRO\ buffer layer can be understood  by an improved surface diffusion. The increased surface diffusion is reminisent to the diffusion enhancement seen in the growth of \SRO\ on \STO\ and for \BFO\ on SrO-terminated {\STO}.\cite{Solmaz2016}$^{,}$\cite{Rijnders2004} The enhanced kinetics subsequently favours the formation of a thermodynamically stable (001) LSMO film while suppressing other orientations or phases (in particular the (110) orientation, seen as a left hand shoulder of the CeO$_2$ (002) reflection), similar to high temperature growth. Apparently, this improved film quality remains over the full thickness of the LSMO film suggesting that this kinetically enhanced growth characteristic is not limited to the interface.

Unlike small scale experiments, for growth on 4'' wafers the plume is scanned and fully collected in its entirety on the wafer. Scanning of the spatially dependent composition of the plume results in a compositional varying flux of species \cite{Groenen2015a}$^{,}$\cite{Orsel2015}. Limited surface diffusion and species intermixing results in mixed phase nucleation and amorphous growth as observed in the structural characterisation results on films without \SRO\ layer at lower temperatures. Note that without scanning, LSMO(001) does form (using similar buffer layers) at 600$^\circ$C, see Fig. S1. \cite{Groenen:supplement} Furthermore, flux changes across a wafer also result in a varying ratio between (110) and (001) oriented LSMO, pointing to the importance of the role of surface kinetics versus flux, see Fig. S2. \cite{Groenen:supplement}

%More specifically, recent experiments show the temporal and spatial dependence of the composition of the plasma in relation to film crystalline characteristics \cite{Groenen2015a}$^{,}$\cite{Orsel2015}. The spatial composition and flux of the plume strongly depends on, for instance, the oxygen pressure, were species, depending on their chemical nature, oxidise when interacting at increased oxygen background gas pressure ($>$0.01mbar). Therefore, the typical oxidation path length as a function of distance and angle with respect to the normal of the target has significant influence on the chemical nature of arriving species and subsequent film stoichiometry. 

%Supplemental measurements in which XRD structural characterisation at multiple positions on the wafer is carried out for LSMO films grown without the plume scanning, collecting the stationary plume fully on the wafer, indicate that at high temperatures, various \LSMO\ phases do form in ratios depending on the position.

%Also addressed in the supplemental data are preliminary supporting measurements in which structural characterisation of films grown at similar temperatures are shown for a stationary plume, where crystalline phases form, while when scanning the plume crystallinity is nearly absent. This improved surface diffusion by elevated temperature or altered surface termination is necessary to overcome nucleation of varying phases as the results of this temporal and spatially dependent composition of the plume. 

%\section{Conclusions}
We have shown results on the growth of LSMO thin films using optimised high quality YsZ // CeO$_2$ // \SRO\ buffer layers on 4'' silicon wafers. The LSMO growth temperature dependence was shown, where with the introduction of a thin \SRO\ layer between CeO$_2$ and LSMO, crystalline and ferromagnetic LSMO was obtained for growth temperatures as low as 250$^\circ$C. When this \SRO\ layer is lacking a fully amorphous or highly polycrystalline film growth is observed at temperatures up to 550$^\circ$C without any ferromagnetic behaviour. We suggest that occurrence of crystal growth of LSMO at these low growth temperatures can be understood by an improved surface diffusion induced with the introduction of an \SRO\ layer. This improved diffusion is necessary to promote species intermixing for correct phase formation. The intermixing apparently is necessary because whilst scanning the full plume with a compositional varying flux of species is collected on the 4'' wafer. Subsequently, it has been shown that by surface diffusion enhancement makes it possible to obtain high quality crystalline growth at lower temperatures. This indicates that the challenges in large scale integration of oxides on silicon and industrial process conditions can be addressed and overcome.

%\begin{acknowledgments}
%
%Financial support is acknowledged from the European Commission - DG research and innovation to the collaborative research project named Interfacing oxides (IFOX, Contract No.  NMP3-LA-2010-246102)
%
%\end{acknowledgments}

%\bibliography{refs}

%

\end{document}